\documentclass{elsarticle}
\usepackage{moreverb}
\usepackage[normalem]{ulem}
\usepackage{subcaption}
\usepackage{cleveref}
\usepackage{gensymb}
\begin{document}

\newcommand{\midrule}{\hline}
\newcommand{\toprule}{\hline}
\newcommand{\bottomrule}{\hline}
\newcommand{\figures}[1]{%
\begin{figure}
\centering
        \begin{subfigure}[t]{0.3\textwidth}
                \includegraphics[trim=60 40 20 40,clip,width=\linewidth]{WPD#1.pdf}
                \caption{WPD at #1 m (original)}
                \label{fig:#1DWH}
        \end{subfigure}%
        ~ %add desired spacing between images, e. g. ~, \quad, \qquad, \hfill etc.
        \begin{subfigure}[t]{0.3\linewidth}
                \includegraphics[trim=60 40 20 40,clip,width=\textwidth]{WPD_#1_i.pdf}
                \caption{WPD at #1 m (ensemble i)}
                \label{fig:#110mH}
        \end{subfigure}
        ~ %add desired spacing between images, e. g. ~, \quad, \qquad, \hfill etc.
        \begin{subfigure}[t]{0.3\textwidth}
                \includegraphics[trim=60 40 20 40,clip,width=\linewidth]{WPD_#1_i_diff.pdf}
                \caption{Relative difference (i-original)}
                \label{fig:#13hrH}
        \end{subfigure}

%        \begin{subfigure}[t]{0.3\textwidth}
% \includegraphics[trim=60 40 20 40,clip,width=\textwidth]{dummy.png}
%        \end{subfigure}%
%        ~ %add desired spacing between images, e. g. ~, \quad, \qquad, \hfill etc.
        \begin{subfigure}[t]{0.3\linewidth}
                \includegraphics[trim=60 40 20 40,clip,width=\textwidth]{WPD_#1_h.pdf}
                 \caption{WPD at #1 m (ensemble h)}
                \label{fig:#110mH}
        \end{subfigure}
        ~ %add desired spacing between images, e. g. ~, \quad, \qquad, \hfill etc.
        \begin{subfigure}[t]{0.3\linewidth}
                \includegraphics[trim=60 40 20 40,clip,width=\textwidth]{WPD_#1_f.pdf}
                \caption{WPD at #1 m (ensemble f)}
                \label{fig:#110mH}
        \end{subfigure}
        ~ %add desired spacing between images, e. g. ~, \quad, \qquad, \hfill etc.
        \begin{subfigure}[t]{0.3\linewidth}
                \includegraphics[trim=60 40 20 40,clip,width=\textwidth]{WPD_#1_g.pdf}
                \caption{WPD at #1 m (ensemble g)}
                \label{fig:#110mH}
        \end{subfigure}

        \begin{subfigure}[t]{0.3\textwidth}
                \includegraphics[trim=60 40 20 40,clip,width=\linewidth]{WPD_#1_h_diff.pdf}
                \caption{Relative difference (h-original)}
                \label{fig:#13hrH}
        \end{subfigure}
        ~ %add desired spacing between images, e. g. ~, \quad, \qquad, \hfill etc.
        \begin{subfigure}[t]{0.3\textwidth}
                \includegraphics[trim=60 40 20 40,clip,width=\linewidth]{WPD_#1_f_diff.pdf}
                \caption{Relative difference (f-original)}
                \label{fig:#13hrH}
        \end{subfigure}
        ~ %add desired spacing between images, e. g. ~, \quad, \qquad, \hfill etc.
        \begin{subfigure}[t]{0.3\textwidth}
                \includegraphics[trim=60 40 20 40,clip,width=\linewidth]{WPD_#1_g_diff.pdf}
                \caption{Relative difference (g-original)}
                \label{fig:#13hrH}
        \end{subfigure}
\caption{Comparison of WPD computations using the ensembles and the original data set. \label{fig:#1_field_err}}
\end{figure}}

\newcommand{\figuresRCP}[1]{%
\begin{figure}
\centering
        \begin{subfigure}[t]{0.3\textwidth}
                \includegraphics[trim=60 40 20 40,clip,width=\linewidth]{WPD#1.pdf}
                \caption{WPD at #1 m (original)}
                \label{fig:#1DWH}
        \end{subfigure}%
        ~ %add desired spacing between images, e. g. ~, \quad, \qquad, \hfill etc.
        \begin{subfigure}[t]{0.3\linewidth}
                \includegraphics[trim=60 40 20 40,clip,width=\textwidth]{WPD_#1_RCP85.pdf}
                \caption{WPD at #1 m (RCP8.5)}
                \label{fig:#110mH}
        \end{subfigure}
        ~ %add desired spacing between images, e. g. ~, \quad, \qquad, \hfill etc.
        \begin{subfigure}[t]{0.3\textwidth}
                \includegraphics[trim=60 40 20 40,clip,width=\linewidth]{WPD_#1_RCP85_diff.pdf}
                \caption{}
                \label{fig:#13hrH}
        \end{subfigure}

\caption{Comparison of WPD computations using the RCP8.5 and the original data set. a) Original data set following \cite{Gross15} , b) RCP8.5 data set , c) relative difference\label{fig:RCP#1}}
\end{figure}}

\begin{frontmatter}

\title{Offshore wind energy climate projection using UPSCALE climate data under the RCP8.5 emission scenario}

\author[cicese]{\corref{cor1}M.~S.~Gross}
\ead{mgross@cicese.mx}

\author[cicese]{V.~Magar}
\ead{vmagar@cicese.mx}

\cortext[cor1]{Corresponding author}

\address[cicese]{Centro de Investigaci\'{o}n Cient\'{i}fica
y de Educaci\'{o}n Superior de Ensenada, Departamento de Oceanograf\'{i}a F\'{i}sica, Carretera Ensenada-Tijuana 3918, Ensenada BC 22860, MEXICO. Tel:+52  646 175 0500 Fax: +52 646 174 4729}

\begin{abstract}

Recently \cite{Gross15} it was demonstrated how climate data can be utilized to estimate regional wind power densities. In particular it was shown that the quality of the global scale estimate compared well with regional high resolution studies and a link between surface temperature and moist density in the estimate was presented. In the present paper the methodology is tested further, to ensure that the results using one climate data set are reliable. This is achieved by extending the study to include four % (xgxqf, xgxqg, xgxqh, xgxqi) further 
ensemble members. With the confidence that one instantiation is sufficient a climate change data set%(xgxqk, xgxql, xgxqm) 
, which was also a result of the UPSCALE\cite{gmd-7-1629-2014,Donlon2012140} experiment, is analyzed. This, for the first time, provides a projection of future changes in wind power resources using this data set. This climate change data set is based on the Representative Concentration Pathways (RCP) 8.5 \cite{RCP,gmdd-4-997-2011,gmd-4-543-2011} climate change scenario. This provides guidance for developers and policy makers to mitigate and adapt.
\end{abstract}


\end{frontmatter}

\section{Introduction}

%IPCC Special Report on
%Renewable Energy Sources
%and
%Climate Change Mitigation:
%
%[7.2]. 

The Working Group III Special Report on Renewable Energy Sources and Climate Change
Mitigation (SRREN) \cite{IPCCSR-Sum} stated that ``research to date suggests that climate change is not expected to greatly impact the global technical potential for wind energy development but changes in the regional 
distribution of the wind energy resource may be expected''.

In \cite{Pryor2010430} the authors provide substantial background information on the use of climate models and their application on the estimation of the impact of climate change on wind energy resources. Whilst mainly focusing on Europe, they find that the changes to be anticipated are below 3\% reduction (next 50 years)
or below 5\% reduction (next 100 years), citing \cite{Breslow2002585}, in the USA. However, \cite{Breslow2002585} use a $2.5\degree$ Hadley Centre Coupled Model (HadCM) II output - contrasted here with the $1/3\degree$ resolution, four times as many vertical levels and the inclusion of density in the estimation. However, uncertainty exists with significant differences between studies. \cite{Breslow2002585} conclude that mean wind speeds may be reduced by 10 to 15\%, and, considering that wind power generation is a function of the cube of the wind speed, such a decrease will correspond to reductions in wind power generation (i.e. revenue of the operator) on the order of 30 to 40\%.

Using statistical down scaling methods, \cite{Sailor20082393} improve the output of the Intergovernmental Panel on Climate Change (IPCC)  global climate models (GCMs) (with a highest resolution of $1.9\degree$). Their results suggest a seasonal component of the climate change impact, with summertime wind speeds in the Northwest USA decreasing by 5-10\%, and low or no impact on winter months. At typical turbine hub heights a 40\% reduction in summertime generation potential is projected. From this work it is clear that higher resolution models have to be used in order to provide suitable, less ambiguous results.

It is important to note here that the terminology in global modeling of the atmosphere is in continuous flux. What used to be the resolution of a regional climate model (RCM) is now the resolution of a GCM. For example \cite{Pryor17052011} state that atmosphere-ocean global climate model (AOGCM) resolution is inappropriate to accurately characterize wind climates and then suggest a RCM model at $0.44 \times 0.44\degree$. That resolution is still coarser than the UPSCALE GCM resolution utilized in this study.

\cite{Pryor05} present results for Europe, analyzing the climate change impact of the A2 scenario using a RCM with (highest) resolution of $0.44\times 0.44\degree$, which is still coarser than the global UPSCALE resolution. It is emphasized that much of the solution in the RCM is dominated by the boundary conditions - another reason to use GCM data.

\cite{GRL:GRL21432} perform empirical down scaling on GCM results with the  finest resolution of $1.875\times 1.875\degree$, for Northern Europe. Down scaling is required due to the coarse temporal and spatial resolution. The A2 emission scenario is analyzed and significant changes have been reported. The A2 scenario equates to a moderate to high greenhouse gas cumulative emission. This results in global carbon dioxide emissions from industry and energy in 2100 that are almost four times the 1900 value \cite{emission_scenario}. The down scaled mean and 90th percentile wind speed over Northern Europe during the 21st century are likely to differ from those that prevailed during the end of the 20th century by less than $\pm 15\%$. They report that this signal is currently comparable to the variation in down scaling results, due to variations in GCM simulation of the down scaling predictors.

%Full report:
%Please use the following reference to the whole report:
%IPCC, 2011: IPCC Special Report on Renewable Energy Sources and Climate Change Mitigation. Prepared by Working Group III of the Intergovernmental Panel on Climate Change [O. Edenhofer, R. Pichs-Madruga, Y. Sokona, K. Seyboth, P. Matschoss, S. Kadner, T. Zwickel, P. Eickemeier, G. Hansen, S. Schlömer, C. von Stechow (eds)]. Cambridge University Press, Cambridge, United Kingdom and New York, NY, USA, 1075 pp.
%
%wind energy chapter:
%Wiser, R., Z. Yang, M. Hand, O. Hohmeyer, D. Infi eld, P. H. Jensen, V. Nikolaev, M. O’Malley, G. Sinden,
%A. Zervos, 2011: Wind Energy. In IPCC Special Report on Renewable Energy Sources and Climate Change
%Mitigation [O. Edenhofer, R. Pichs-Madruga, Y. Sokona, K. Seyboth, P. Matschoss, S. Kadner, T. Zwickel,
%P. Eickemeier, G. Hansen, S. Schlömer, C. von Stechow (eds)], Cambridge University Press, Cambridge, United
%Kingdom and New York, NY, USA

\section{The climate data set}

The data set has been described in \cite{Gross15}. In the context of the present work only the details of the climate change run need to be detailed. The climate change simulations were configured with sea surface temperature (SST)s comprised of the Operational Sea Surface Temperature and Sea Ice Analysis (OSTIA) data \cite{Donlon2012140}, as used in the present climate runs, plus the SST change between 2000 and 2100 in the Hadley Centre Global Environmental Model version 2 Earth System (HadGEM2-ES) runs under the RCP8.5 climate change scenario (cf. \cite{gmdd-4-997-2011,gmd-4-543-2011}).

\section{Results}

\begin{figure}
\centering
\includegraphics[width=\linewidth]{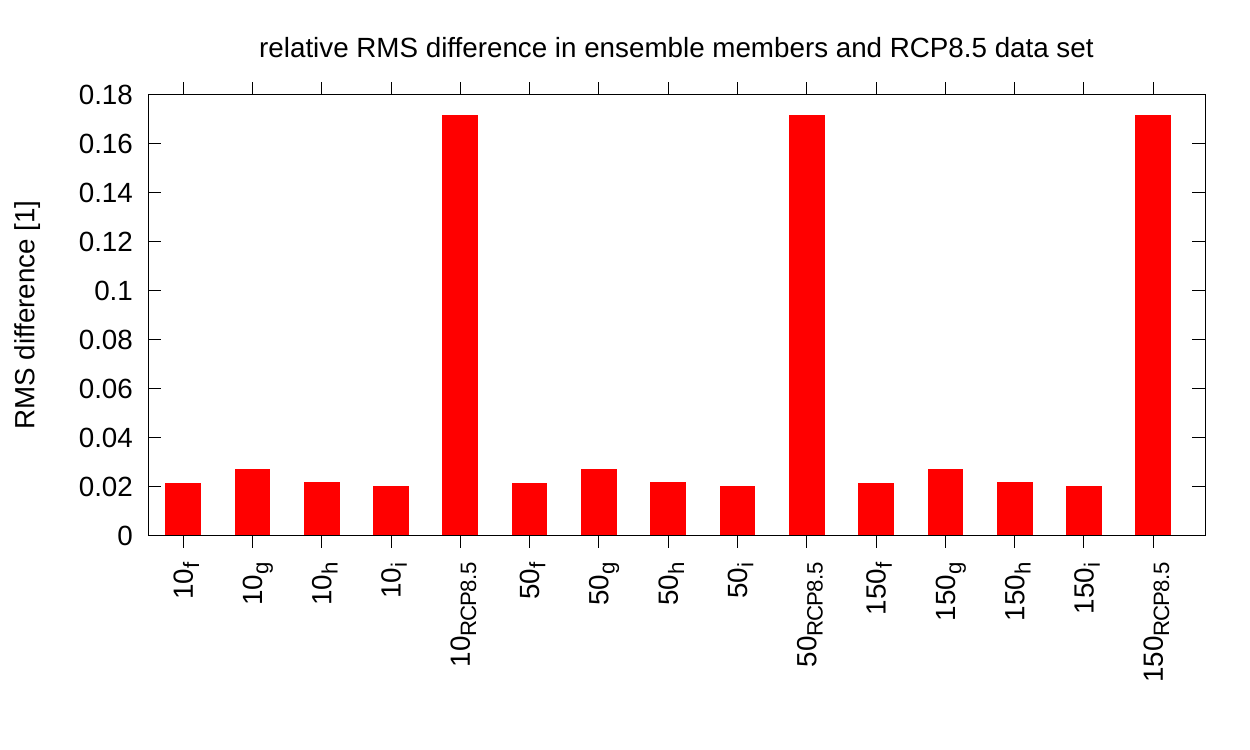}
\caption{ Relative RMS difference for ensembles and climate change data set\label{fig:RMS}}
\end{figure}

The relative difference has been computed according to 
\begin{equation}
\delta_{rel}=\frac{WPD_{ens}-WPD_{ref}}{WPD_{ref}}.
\end{equation}
for each of the ensemble members $WPD_{ens}$, using the data set from \cite{Gross15} as a reference $WPD_{ref}$.
Representative field plots for a hub height of $50$m are shown in Figure \ref{fig:50_field_err}.
%\figures{10}
\figures{50}
%\figures{150}
The respective plots for $10$m and $150$m hub heights are omitted because they differ only slightly.
Then the relative root mean square (RMS) difference, $RMS_{rel}$ 
\begin{equation}
 RMS_{rel}=\sqrt{\frac{1}{n} \displaystyle\sum{\delta_{rel}^{2}}}
\end{equation}
can be computed.
The results are shown in Figure \ref{fig:RMS}. The respective RMS differences are labeled according to the hub height and the ensemble member, i.e. $10_{f}$ represents the difference for the $10$m hub height, ensemble member $f$. $10_{RCP8.5}$ similarly represents the  $10$m hub height and the climate change run.

\subsection{Extended sets}
\begin{figure}
\centering
\includegraphics[width=\linewidth]{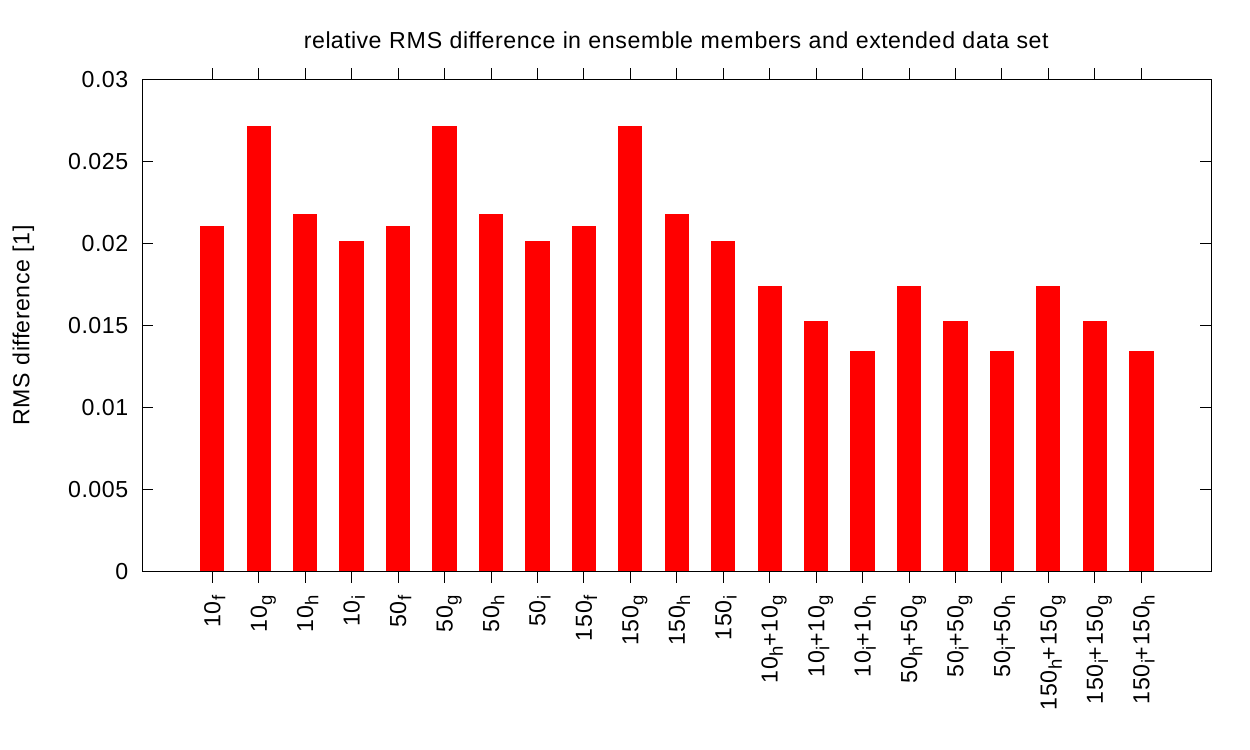}
\caption{ Relative RMS difference for ensembles and extended data set\label{fig:RMS_ext}. The non extended data sets are compared to the original data set and the extended data sets are compared to the original data set extended by ensemble member $f$.}
\end{figure}

The perturbed ensemble member can be seen as an independent set of measurements. This means that also any combination of ensemble members is a new set of measurements. Therefore the ensemble members can be combined. Using these extended sets the effect of a data set of twice the length or double the sampling rate can be investigated.  This is possible because instantaneous measurements, not the $3$hr mean are used. Using the original data set augmented with ensemble member $f$ as control $(WPD_{ref})$, three unique data points can be computed using the $g+h$, $g+i$ and $h+i$ combinations. The RMS difference, for example
\begin{equation}
 RMS_{rel}=\sqrt{\frac{1}{n} \displaystyle\sum{\left(\frac{WPD_{g+h}-WPD_{ref}}{WPD_{ref}}\right)^{2}}},
\end{equation}
for these three sets is lower (although not by much) than the one for the original sets, as expected, due to more data being used. See Figure \ref{fig:RMS_ext} for a comparison.
% For three data points and a confidence interval of $95\%$ the critical $W$ value is $0.767$. The results of the Shapiro Wilks test \cite{SHAPIRO01121965} are provided in Table \ref{tab:Shapiro2}.
%\begin{table}
%\caption{\label{tab:Shapiro2} Shapiro-Wilks test results and standard deviation of the RMS differences of the constructed (extended) sets. }
%\centering
%\small
%\begin{tabular}{lccc}
%\toprule
%height &  $W$  & $p_{SW}$ & $2\sigma$  \tabularnewline
%\midrule
%10      & 0.99751 &0.9047 & 0.00325 \tabularnewline
%50      & 0.99750 &0.9045 & 0.00325 \tabularnewline
%150     & 0.99747 &0.77724& 0.00325 \tabularnewline
%\bottomrule
%\end{tabular}
%\end{table}

Comparing the two sets, the original and extended, using the Mann-Whitney-Wilcoxon (MWW) Ranksum test \cite{Wilcoxon1945,mann1947} yields $p_{MWW}$ values of $0.03389$, $0.03389$ and $0.04953$ for the $10$m, $50$m and $150$m results, respectively.  With the $p_{MWW}$ values $\le 0.05$ it therefore can be claimed that the distributions differ significantly (albeit just), and the extension had a statistically significant, albeit small, impact on the difference. Comparing the extended sets with the original data set yields a RMS difference of $0.0228>RMS_{rel}>0.0188$, i.e. in a similar range as the ensemble members (the first 12, $10_{f}$ to $150_{i}$, in Figure \ref{fig:RMS_ext}). At this level it is likely that other factors in the methodology are more important, such at the extrapolation, resolution and parameterizations used in the GCM.

\subsection{The climate change data set}
Having shown that the data from one run already provides good guidance (i.e. the benefit of using more data is small/negligible), the focus can now shift to the analysis of the climate change data. It can be seen that the difference between the RCP8.5 data set is substantial (Figure \ref{fig:RMS}), when compared with the internal (natural) variability of the climate system, represented in the RMS difference in between the ensemble members. The $W$ values of the Shapiro-Wilks test \cite{SHAPIRO01121965} (see Table \ref{tab:Shapiro})  indicate a normal distribution of the RMS differences of the original data set and the ensemble members. With a 95\% confidence interval the critical $W$ value is $0.748$. Here the $W$ values are greater than the critical $W$ and $p_{SW}>\alpha = 0.05$. Therefore the data is likely normal distributed. Assuming it is normal distributed, a deviation of $2\sigma$ is statistically relevant and hence the difference of the climate change run difference (of $\approx 25\sigma$) is statistically relevant. The field plots of the difference, RCP8.5 climate change data set versus original dataset, is reproduced in Figures \ref{fig:RCP10}, \ref{fig:RCP50} and \ref{fig:RCP150}.

\begin{table}
\caption{\label{tab:Shapiro} Shapiro-Wilks test results and standard deviation of the RMS differences. }
\centering
\small
\begin{tabular}{lccc}
\toprule
height &  $W$  & $p_{SW}$ & $2\sigma$  \tabularnewline
\midrule
10      & 0.815 &0.132792 & 0.0054   \tabularnewline
50      & 0.815 &0.132898 & 0.0054   \tabularnewline
150     & 0.845 &0.227953& 0.0062   \tabularnewline
\bottomrule
\end{tabular}
\end{table}

\figuresRCP{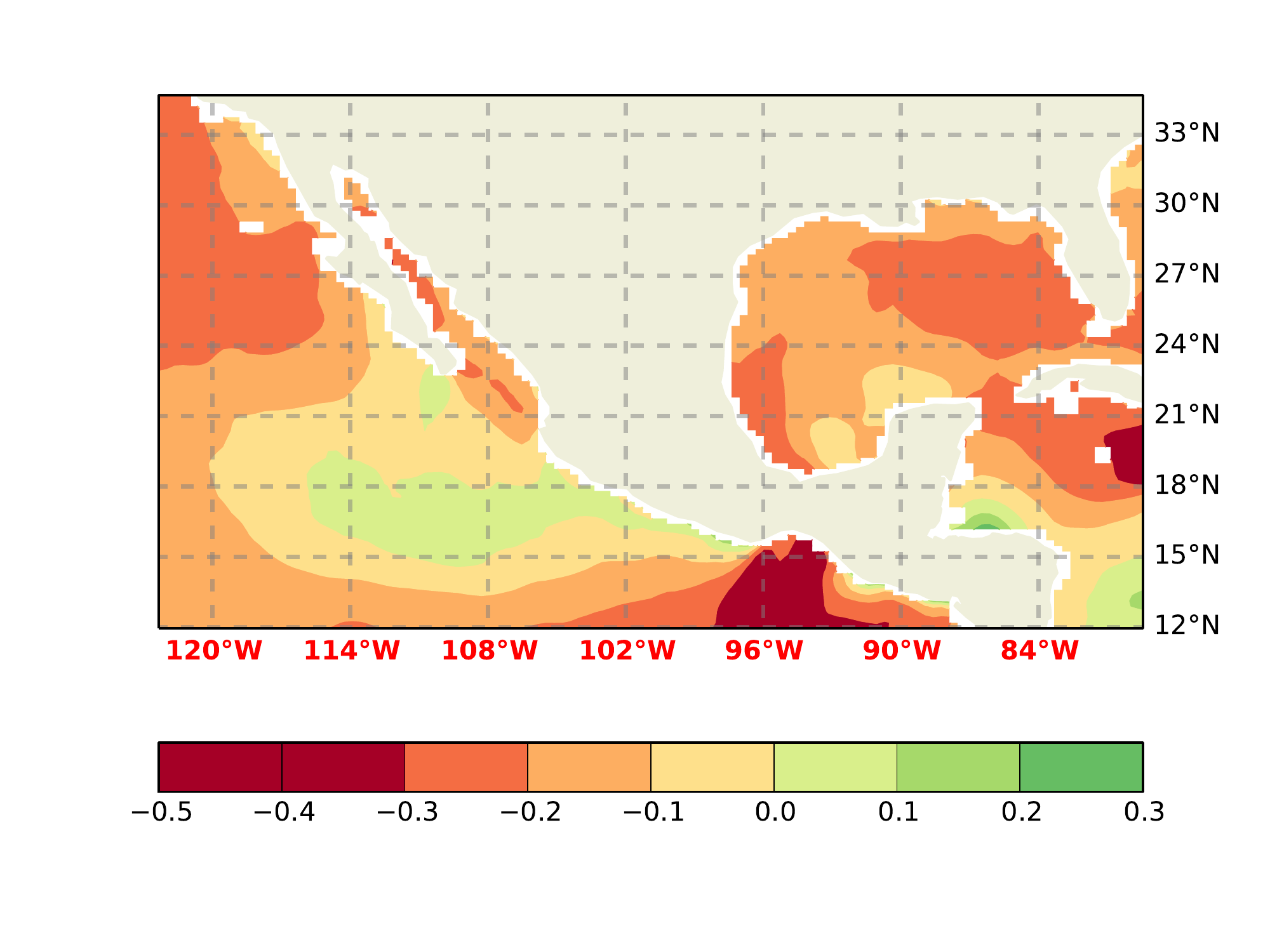}
\figuresRCP{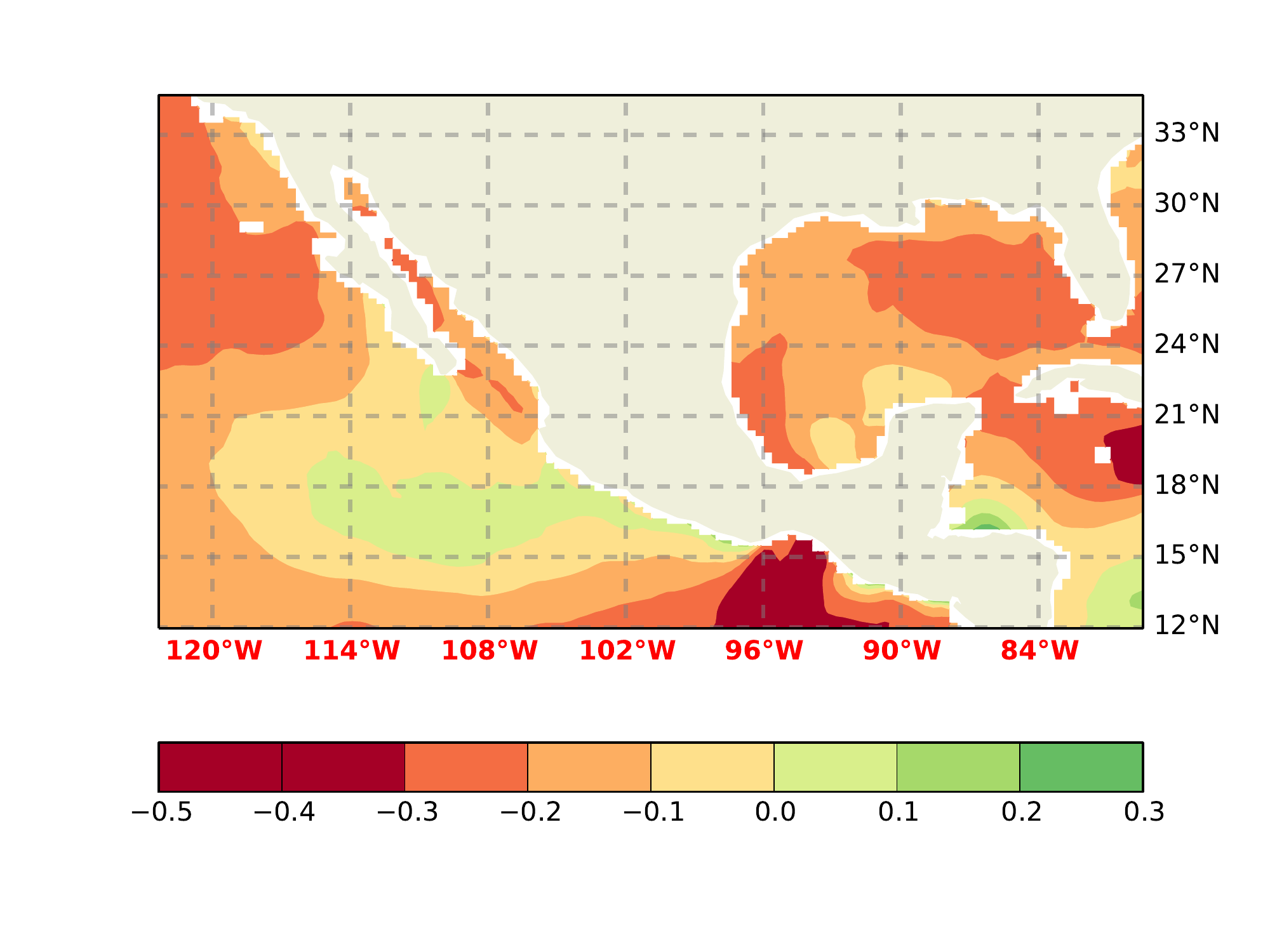}
\figuresRCP{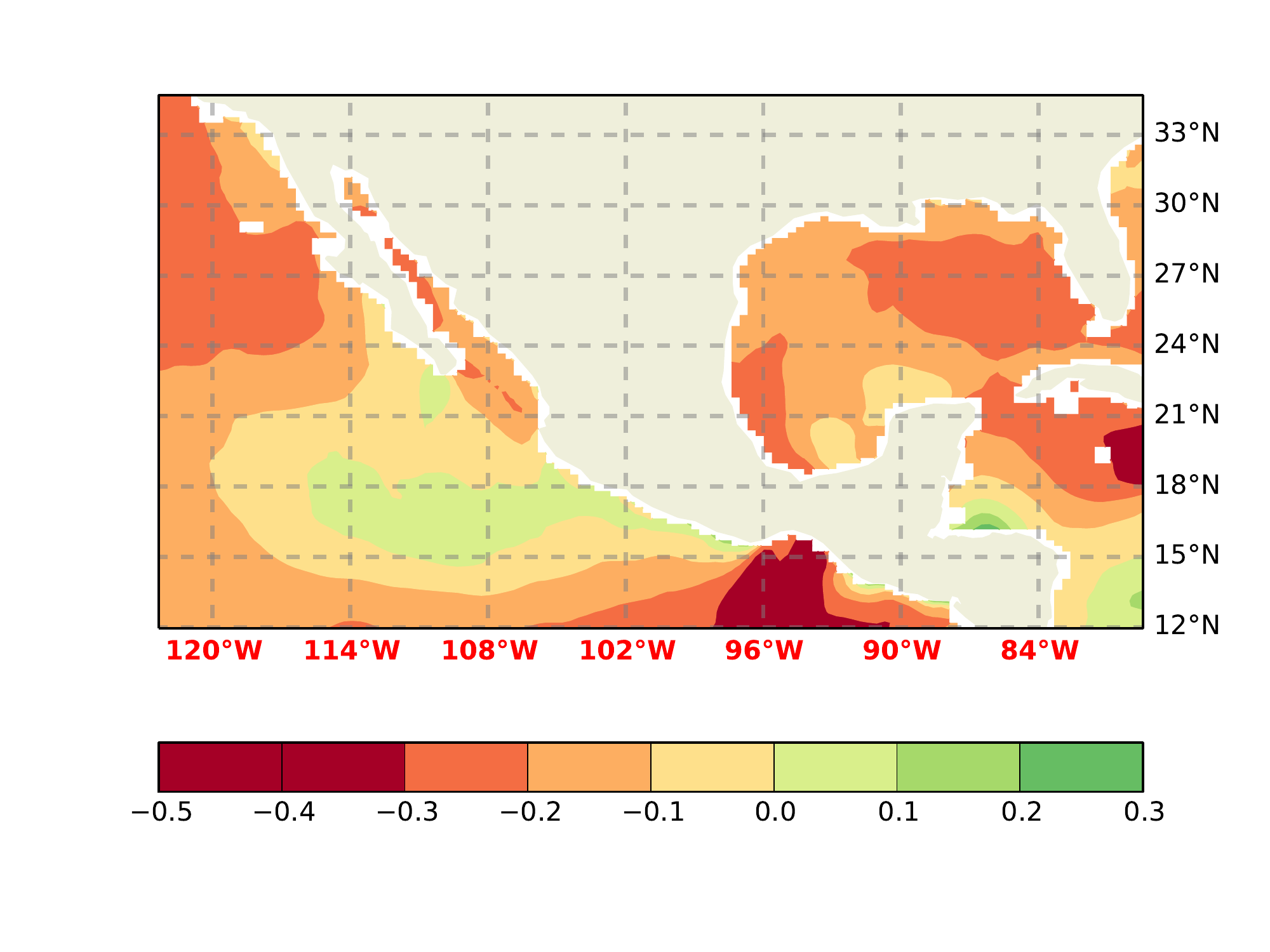}

\subsection{Regional results}
\begin{figure}
\centering
                \includegraphics[width=\linewidth]{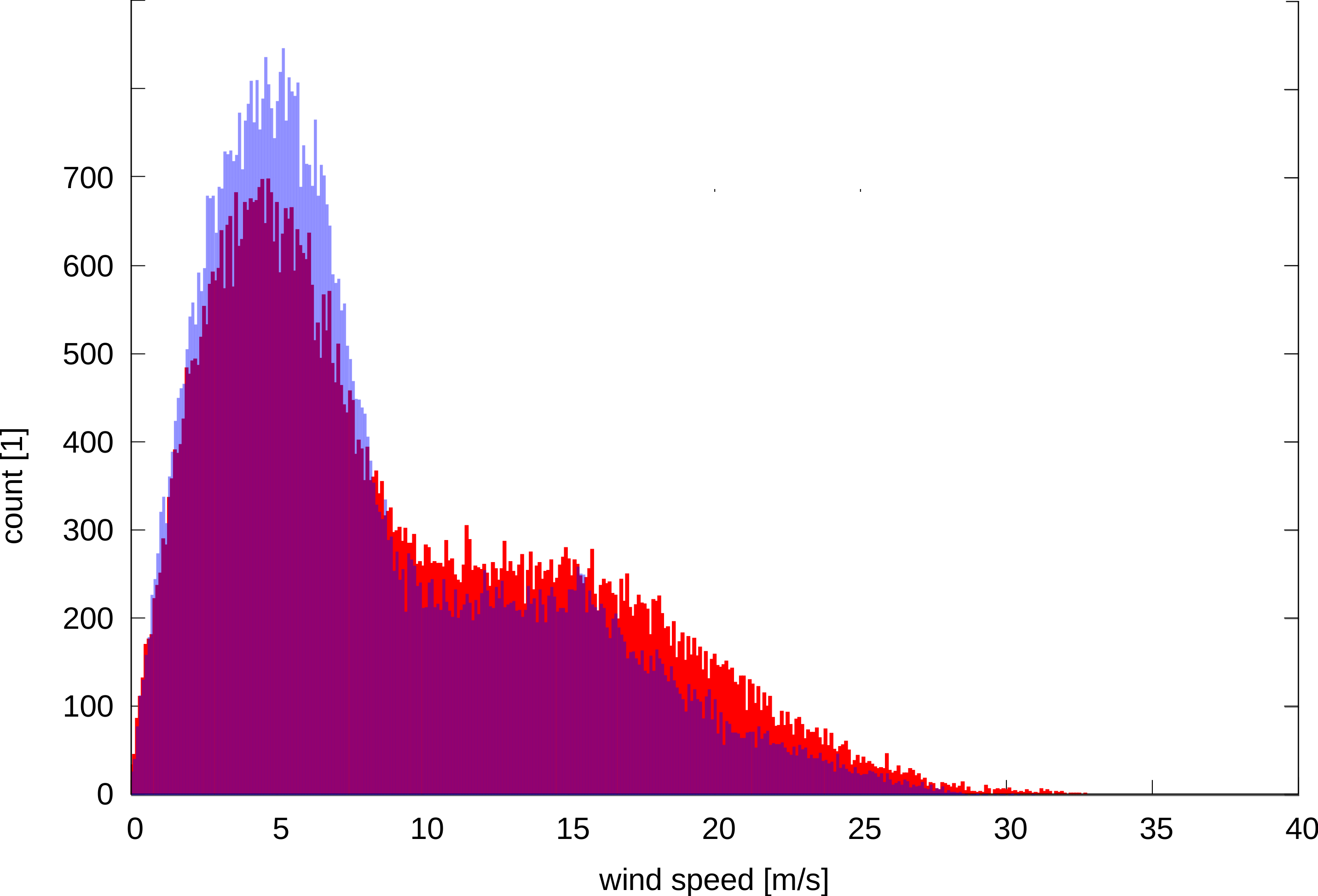}
                \caption{Wind speed histogram for the Gulf of Tehuantepec (latitude = 15.2344, longitude = 265.078). Present climate in red and RCP8.5 climate in translucent blue.}
                \label{fig:hist-Tehuantepec}
\end{figure}

Whilst the regional results are only of relevance to the developers and policy makers in the respective region there is still general interest in what information can be obtained from the data set (which is global). Only as a representative example, the wind speed histogram for the Gulf of Tehuantepec is shown here in Figure \ref{fig:hist-Tehuantepec}, where the climate change signal indicates a reduction of 50\% in the WPD. The present climate data set is red and the RCP8.5 data set in translucent blue.
It is interesting to note that there is a \emph{reduction} in high wind events, and an increase in average wind speed events. This is both, good news to the developers and contrary to the general assumption that with climate change extreme events will occur more frequently.

\section{Conclusion}

Here it was shown that the altered climate of a high resolution GCM simulation does have a marked impact on the projected wind power density. As far as the climate results can be relied on, and there obviously is still a lot of discussion and ongoing work in the scientific community, this will provide important insights in the long term scenarios than can be expected.

It was also shown that using one time series of 74880 samples already produces a reasonably solid estimate. Doubling the sample size did improve the results, in a statistically significant way, however, only a modest difference was observed, certainly in contrast to the response to the climate change run.

From an environmental perspective it is disappointing to see the broad scale drop in projected average wind power densities. However, sensible technical and financial decisions should be able to incorporate or mitigate this. In any case, as well as the uncertainties underlying this analysis (small size, resolution, extrapolation etc) it is not given that the future climate will exactly instantiate RCP8.5.

Technologically this means that current installations have to be efficient at the current and (significantly) lower than current wind speed levels. 

Financial decisions have to take the anticipated decline in output (where this happens to be the case) into consideration, and this decline in WPD increases the risk rating for sites which currently evaluate as having marginal potential for development.

\section*{Acknowledgments}
This work is based on the UPSCALE data set licensed from the University of Reading
which includes material from Natural Environment Research Council (NERC) and the Controller of Her Majesty’s Stationery Office (HMSO) \& Queen’s
Printer. The UPSCALE data set was created by P. L. Vidale, M. Roberts, M.
Mizielinski, J. Strachan, M.E. Demory and R. Schiemann using the HadGEM3
model with support from NERC and the Met Office and the PRACE Research
Infrastructure resource HERMIT based in Germany at High Performance Computing Center Stuttgart (HLRS).

Use of the MONSooN system, a collaborative facility supplied under the Joint Weather and Climate Research Programme, a strategic partnership between the Met Office and the NERC, is acknowledged.

\bibliographystyle{elsarticle-num}
\bibliography{refs}

\end{document}